\documentstyle[referee,psfig]{l-aa}
\begin{document}

\newcommand{\bdouble}{\baselineskip 2.0\baselineskip}         

   \thesaurus{06     
              (02.01.2;  
               08.02.1;  
               08.09.2;  
               08.14.1;  
               13.25.5)}  
   \title{GX\,349+2 (Sco\,X-2): An odd-ball among the Z sources}

   \author{Erik Kuulkers
           \inst{1}
	   \and
	   Michiel van der Klis
	   \inst{2}
          }

   \offprints{Erik Kuulkers}

   \institute{Astrophysics, University of Oxford, Nuclear and Astrophysics 
	      Laboratory, Keble Road, Oxford OX1 3RH, United Kingdom\\
              email: erik@astro.ox.ac.uk
	  \and
	      Astronomical Institute ``Anton Pannekoek'',
              University of Amsterdam and Center for High-Energy Astrophysics,
              Kruislaan 403, NL-1098 SJ Amsterdam, The Netherlands\\
	      email: michiel@astro.uva.nl
             }

   \date{Accepted by A\&A}

   \maketitle


   \begin{abstract}

We report on about 4\,hrs of observations made with the Rossi X-ray Timing Explorer
on 1997 May 2 of the low-mass X-ray binary and Z source GX\,349+2 (Sco\,X-2).
Initially the source was in the normal branch (NB), later it moved to the
flaring branch (FB). In the NB the power spectra reveal a broad 
(FWHM$\sim$16\,Hz) noise component peaking near 9\,Hz, with a fractional 
rms amplitude of
$\sim$3\%\ (2--30\,keV). This noise component does not resemble the strong
quasi-periodic oscillations (QPO) usually seen in the NB of other Z sources. We
set 95\%\ confidence upper limits on the fractional rms of $\sim$0.9\%\
(2--30\,keV) on any such QPO. In the FB the power spectrum showed a 
somewhat less broad noise component (FWHM$\sim$11\,Hz) peaking near 6\,Hz with a
fractional rms of $\sim$4\%\ (2--30\,keV). We compare our results with previous
reports, and find that the fast timing behaviour changes not only with
position in the Z, but also as 
a function of the position of the Z track in the hardness-intensity diagram.
By comparing GX\,349+2 with the other Z and atoll sources, we conclude that 
GX\,349+2 differs from the Z sources in various aspects and shows
similarities to the behaviour seen in the bright atoll sources, such as 
GX\,13+1 and GX\,3+1.
We also searched for kilo-Hertz QPO, similar to those present in other Z
sources. We only found weak evidence (2.6\,$\sigma$ confidence) for a QPO near
1020\,Hz with FWHM$\sim$40\,Hz and fractional rms of $\sim$1\%. We note that 
this frequency and fractional rms are consistent with those expected and 
observed in the lower NB/FB of the Z sources Sco\,X-1, GX\,17+2 and Cyg\,X-2.

   \end{abstract}


\section{Introduction}

The group of the Z sources consists of 6 bright low-mass X-ray binaries
(LMXBs): Sco\,X-1, GX\,5$-$1, GX\,349+2 (or Sco\,X-2), GX\,17+2, 
GX\,340+0 and Cyg\,X-2 (Hasinger \&\ van der Klis \cite{hk89}, van der Klis
\cite{k95book}). 
GX\,349+2 was classified as a Z source based on its narrow branches in 
the X-ray color-color diagram (CD) and its steep very-low-frequency 
($\la$1\,Hz) noise slope. The Z sources trace out Z-like shaped tracks in the 
CD, with the limbs of the Z called horizontal branch (HB), normal 
branch (NB) and flaring branch (FB). It is thought that 
mass accretion rate, \.M, increases from the HB, through the NB, to the FB. 
All the six sources have displayed the three branches, 
except for GX\,349+2 which has up to now only been seen
in the NB and FB (see Hasinger \&\ van der Klis \cite{hk89}; 
Kuulkers \&\ van der Klis \cite{kk95a}).

The fast time variability is closely related to position in the Z.
In the HB and upper part of the NB quasi-periodic oscillations (QPO) are 
present with frequencies between $\sim$15--60\,Hz (HBO) together with a noise 
component below $\la$20\,Hz (called low-frequency noise, LFN). On the NB a 
different QPO (NBO) is present with frequencies between 5--7\,Hz, FWHM 
between 2--4\,Hz and fractional rms of 1--4\%\ ($\sim$2--10\,keV). 
On the lower part of the FB in Sco\,X-1 and GX\,17+2 the NBO merge smoothly 
into FB QPO (FBO) with frequencies of up to $\sim$20\,Hz.

All Z sources show shifts in the position of the Z track in the 
hardness-intensity diagram (HID). Only in Cyg\,X-2, GX\,5$-$1, and GX\,340+0 
have these been observed to be accompanied by Z-track shifts and morphology 
changes in the CD (see Kuulkers \&\ van der Klis 1995a). In the case of 
GX\,349+2, two EXOSAT observations showed two tracks shifted in intensity in 
the HID, and we therefore refer to these two tracks as the Z observed at 
``low overall intensities'' and ``high overall intensities''.

Recently, the Rossi X-ray Timing Explorer (RXTE) has proven to be a
gold mine for the detection of kHz QPO in LMXBs (e.g.\ van der Klis \cite{k97}, 
and references therein). So far, kHz QPO in Z sources have been found
in Sco\,X-1, GX\,5$-$1, GX\,17+2 and Cyg\,X-2 
(van der Klis et al.\ 1996a,b, 1997a,b, Wijnands et al.\ 1997b,c). 
The frequency of the kHz QPO increases with increasing \.M.

GX\,349+2 is a poorly studied Z source. Therefore, much of the properties 
inferred from the other Z sources have been assumed to apply also for GX\,349+2.
In this paper we report on observations of GX\,349+2 as observed with the 
RXTE PCA, compare them with previous observations, and show that 
the fast time variability is somewhat different from that observed in the 
other Z sources.

\section{Observations and Analysis}

The RXTE PCA (Bradt et al.\ \cite{brs93}) observed GX\,349+2 on 1997 May 2 
18:44--22:54~UTC. The data were collected from all five proportional
counter units (PCUs) with a time 
resolution of 16\,s (129 photon energy channels, effectively covering 
2.0--30\,keV) and 125\,$\mu$s (single-bit mode data in 4 energy channels, 
effectively covering 2.0--5.0--6.4--8.6--30\,keV). 

The intensity used in this paper is defined as the 5-PCU count rate in the 
energy band 2.0--19.7\,keV. For the hardness-intensity diagram (HID) we used 
the 16-s data; the hard color is defined as the count rate ratio between 
8.6--19.7\,keV and 6.0--8.6\,keV. X-ray spectral fits were performed in the 
range 2--30\,keV and a 2\%\ uncertainty was included in the data to account 
for uncertainties in the PCA response matrix (see e.g.\ Cui et al.\ 
\cite{chr97}). The X-ray spectra were corrected for background and dead time.

Power density spectra were made from the single-bit mode data, binned to
250\,$\mu$s, using 16\,s data stretches. 
In order to study the low-frequency ($\la$100\,Hz) behaviour 
we fitted the 0.125--128\,Hz power spectra with a constant representing the
Poisson noise, a Lorentzian and/or a power-law with exponential
cut-off to describe peaked noise components, and a power-law describing 
the underlying continuum (VLFN). To investigate the power spectrum for 
kHz QPO we fitted the 256--2048\,Hz power spectra with a function described 
by a constant, a Lorentzian to describe QPO, and a broad sinusoid to 
represent the dead-time modified Poisson noise  (Zhang et al.\ \cite{zjs95}; 
Zhang \cite{z95}). Errors quoted for the power spectral parameters were 
determined using $\Delta\chi^2$=1. Upper limits were determined using 
$\Delta\chi^2$=2.71, corresponding to 95\%\ confidence levels.

\section{Results}

In Figs.~1a and b we show the light curve and the corresponding HID. 
During the first two contiguous data segments the source was in the 
NB, while in the last one it was in the FB. Since the flaring in the FB 
of GX\,349+2 is known to have higher amplitudes than seen here (e.g.\
Ponman et al.\ \cite{pcs88}), we reckon that 
the source traced out only the lower part of its FB.

We combined all power spectra of the first two segments (NB) and of the third 
segment (lower FB). The noise component between $\sim$3--20\,Hz in the NB 
power spectrum  (Fig.~2a) can be described by a cut-off power-law 
($\chi^2_{\nu}$/degrees of freedom [dof] of 116/65) with a strength of
1.9$\pm$0.3\%\ (0.1--100\,Hz), and power law index and cut-off frequency of 
$-$2.4$\pm$0.4 and 3.8$\pm$0.6\,Hz, respectively. A Lorentzian fit 
to this noise component resulted in a fractional rms, FWHM, and centroid 
frequency, $\nu_{\rm c}$, of 2.7$\pm$0.1\%, 16$\pm$2, and 9.4$\pm$0.5, 
respectively ($\chi^2_{\nu}$/dof of 111/65). Formally this "peaked noise" 
component in the NB is no QPO, since FWHM$>$0.5$\nu_{\rm c}$ 
(e.g.\ van der Klis \cite{k95}).

A cut-off power-law fit to the noise component in the FB power spectrum
gave a fractional rms (0.1--100\,Hz), power-law index and cut-off frequency
of 3.1$\pm$0.3\%, $-$1.9$\pm$0.3, and 3.6$\pm$0.4\,Hz, respectively
($\chi^2_{\nu}$/dof = 70.5/65). A Lorenztian fit to this noise component
resulted in a fractional rms, FWHM, and frequency of 4.2$\pm$0.1\%, 
11.0$\pm$0.7\,Hz, and 5.8$\pm$0.2\,Hz, respectively 
($\chi^2_{\nu}$/dof = 67.3/65). Formally, also this "peaked noise" component 
may not be regarded as QPO. 

In order to study the energy dependence of the fractional rms of the 
peaked noise in the NB and lower FB, we fitted the power spectra in the four 
available energy bands. The peaked noise in the NB was modeled with a cut-off 
power law and that in the FB as a Lorentzian. The power-law index, $\Gamma$, 
and cut-off frequency, $\nu_{\rm cut-off}$, and the centroid frequency, 
$\nu_{\rm c}$, and FWHM were fixed in this process, respectively for the NB 
and FB peaked noise component. The results are given in Table 1. 
The strength of the peaked noise component in the FB increases significantly 
with energy, while that in the NB may do so too, but is consistent with
being constant.

\begin{table}
\caption{NB and FB peaked noise energy dependence}
\begin{tabular}{ccc}
\hline
Effective & NB peaked noise & FB peaked noise \\
energy range & fractional rms amplitude$^1$ & fractional rms amplitude$^2$ \\
(keV) & (\%\/) & (\%\/) \\
\hline
2.0--5.0 & 1.6$\pm$0.1 & 3.1$\pm$0.1 \\
5.0--6.4 & 1.6$\pm$0.3 & 4.3$\pm$0.2 \\
6.4--8.6 & 2.1$\pm$0.3 & 5.0$\pm$0.3 \\
8.6--30  & 2.5$\pm$0.2 & 5.6$\pm$0.2 \\
\hline
\multicolumn{3}{l}{\footnotesize $^1$\,Modeled as cut-off power law; $\Gamma$ and $\nu_{\rm cut-off}$ fixed, see text.} \\
\multicolumn{3}{l}{\footnotesize $^2$\,Modeled as a Lorentzian; FWHM and $\nu_{\rm c}$ fixed, see text.} \\
\end{tabular}
\end{table}

We found no narrow QPO on the NB. We derived 95\%\ confidence 
upper limits on the strength (2.0--30\,keV) of such NBO by adding a 
Lorentzian to the NB power spectrum, fixing the FWHM to a typical value of 
3\,Hz (see e.g.\ van der Klis 1995b). This resulted in upper limits between 
0.6 and 0.9\%, for frequencies between 4 and 9\,Hz.

We also searched for the presence of kHz QPO in GX\,349+2. No strong 
kHz QPO can be found in the power spectra. A weak QPO peak near 1020\,Hz 
was discerned in the 2.0--30\,keV band power spectrum (Fig.~3), however, 
perhaps due to the relative shortness of the observation; it was only 
significant at the 2.6$\sigma$ level (estimated from an F-test for
the inclusion of the QPO and from the 68\%\ confidence error-scan of the
integral power in the $\chi^2$-space, i.e.\ $\Delta\chi^2$=1). 
When described with a Lorentzian the QPO had a 
centroid frequency, FWHM, and fractional rms of
1020$\pm$11\,Hz, 40$\pm$25\,Hz, and 1.0$\pm$0.2\%, respectively.
Since kHz QPO in Z sources are hard (van der Klis et al.\ 1996a,b, 1997a,
Wijnands et al.\ 1997b,c) we determined 
the rms dependence of the 1020\,Hz feature as a function of energy (fixing 
the centroid frequency and FWHM to the above derived values).
This resulted in 95\%\ confidence upper limits 
of 1.6\%, 2.1\%, 2.1\%, and 1.4\%, for the effective energy ranges
2.0--5.0\,keV, 5.0--6.4\,keV, 6.4--8.6\,keV, and 8.6--30\,keV, respectively.
This provides no additional indications as to the reality of the kHz QPO peak.

In order to compare our results with previous EXOSAT observations, we 
determined the position of the source during the RXTE observations 
in the CD and HID of Kuulkers \&\ van der Klis (\cite{kk95a}). We therefore 
fitted a mean NB X-ray spectrum and folded the fit through the EXOSAT 
ME response. The resulting mean value for the intensity corresponds to the 
NB of the EXOSAT HID at the highest overall intensities.

\section{Discussion}

We found peaked noise (Lorentzian with FWHM$\sim$11\,Hz and 
$\nu_{\rm c}$$\sim$6\,Hz) when
GX\,349+2 was in the lower part of the FB with a fractional rms of
$\sim$4\%\ (2--30\,keV). Its strength increased significantly from 
$\sim$3\%\ to $\sim$5.5\%\ with increasing photon energy of $\sim$3\,keV to 
$\sim$10\,keV. In the NB we found a weaker and 
somewhat broader noise component (Lorentzian with rms of $\sim$3\%\ 
[2--30\,keV], FWHM of $\sim$16\,Hz and $\nu_{\rm c}$ of 9.4\,Hz).

Very similar behaviour was reported from extensive 
EXOSAT observations of GX\,349+2 by Ponman et al.\ (\cite{pcs88}). 
Their peaked noise (referred to as ``broad QPO'') was strongest
when GX\,349+2 was in transition between its quiescent (i.e.\ NB) and
peak (upper FB) intensities. When strongest ($\sim$3.2\%, 1--10\,keV) its FWHM
and $\nu_{\rm c}$ were similar to ours (see also Hasinger \&\ van der Klis
1989).
It seems therefore likely we have observed 
the same phenomenon. Our analysis of the RXTE X-ray spectrum in the NB 
shows that indeed our observation falls in the same part of the HID diagram as
the observations by Ponman et al.\ (\cite{pcs88}); see 
Kuulkers \&\ van der Klis (\cite{kk95a}). 
The peaked noise observed by Ponman et al.\ (\cite{pcs88}) became broader in 
the NB (and therefore more consistent with a broad-band noise; see
also Hasinger \&\ van der Klis \cite{hk89}), a similar behaviour as we found. 
Further up the FB the peaked noise became narrower and rapidly weaker
(Ponman et al.\ 1988).

EXOSAT observations taken $\sim$1~year before those described by 
Ponman et al.\ (\cite{pcs88}), however, showed a different behaviour.
At that time the power spectrum showed a peak with fractional 
rms of $\sim$2\%, and centroid frequency and FWHM of $\sim$11\,Hz and 
$\sim$7.5, respectively (Lewin et al.\ \cite{lpj85}), and it was at lower 
overall intensities (see Kuulkers \&\ van der Klis \cite{kk95a}). Observations 
with the Einstein solid state spectrometer (SSS) also revealed a peak with 
centroid frequency of $\sim$11\,Hz and FWHM of $\sim$4\,Hz 
(Christian et al.\ \cite{csk88}), when the source was flaring 
(Christian et al.\ \cite{csk88}; see also Christian \&\ Swank \cite{cs97}). 
Note that this particular peak was narrower than 0.5$\nu_{\rm c}$. The power 
spectral peaks of Lewin et al.\ (1985) and Christian et al.\ (1988)
were reported to be significant at the $\sim$3$\sigma$ level.
We note that folding the ``low-state'' (i.e.\ NB) Einstein SSS 
and monitor proportional counter spectrum of GX\,349+2 (Christian \&\ Swank
\cite{cs97}) through the EXOSAT ME response matrix gives intensities which lie 
in between the mean NB intensities of the two EXOSAT observations in 
Kuulkers \&\ van der Klis (1995).

The above described behaviour is in contrast with that found in Sco\,X-1
(Hertz et al.\ \cite{hvw92}, Dieters \&\ van der Klis \cite{dk97})
and GX\,17+2 (Penninx et al.\ \cite{plm90}). These sources and 
GX\,349+2 show comparable properties in the CD and HIDs (although no
HB has been observed for GX\,349+2) and similar light curves 
(e.g.\ Schulz et al.\ \cite{sht89}, Hasinger \&\ van der Klis 1989).
However, in both Sco\,X-1 and GX\,17+2 a strong NBO 
1--3\%\ rms [$\sim$2--10\,keV], 2--6\,Hz FWHM, 5--7\,Hz centroid frequency) 
merges smoothly into FBO (1--3\%, $\sim$2--10\,keV) when the source moves 
from the NB to the FB. In the FB the FBO frequency increases from 
$\sim$6--7\,Hz up to $\sim$20\,Hz and becomes simultaneously broader 
(up to $\sim$20\,Hz). On the other hand, we note that the energy 
(Ponman et al.\ \cite{pcs88}, this paper) and time lag 
(Ponman et al.\ (\cite{pcs88}) behaviour of the peaked noise in the FB of 
GX\,349+2 are very similar to that seen in the NBO/FBO in Sco\,X-1 
(Dieters et al.\ \cite{dvk97}). In both sources the (broad) QPO increases 
in strength at higher energies and there is no evidence for a lag between 
low-energy and hard-energy QPO. Also in GX\,17+2 the FBO
becomes stronger as a function of energy (Penninx et al.\ \cite{plm90}). 
So, perhaps the peaked noise seen in GX\,349+2 still have a similar 
origin as the NB/FB QPO in Sco\,X-1 and GX\,17+2.

No narrow NBO like that found in the other 5
Z sources (see e.g.\ Hasinger \&\ van der Klis 1989) has ever been observed
in GX\,349+2. We derived upper limits on the presence of typical NBO 
($\sim$0.9\%, 2--30\,keV), i.e.\ below the NBO fractional rms range (1--4\%,
$\sim$2--10\,keV) observed in the other sources.
Moreover, as shown above, the power spectral behaviour of GX\,349+2 at 
roughly the same position in the Z track seems to be different at different 
overall intensities. Changes in the fast timing behaviour in Cyg\,X-2 at the 
same position in the CD or HID as a function of overall intensity were 
recently found by Wijnands et al.\ (\cite{wkk97}). They reported a decrease 
in strength of the NBO from medium (2--3\%\/) to high overall intensities 
($<$1\%\/). Changes in the NBO properties between the different Z sources 
have been discussed by Kuulkers \&\ van der Klis (\cite{kk95b}). NBO were 
suspected to be weak or absent when a source is viewed at a relatively high 
inclination. The absence of a strong NBO in GX\,349+2, however, is in
contradiction with our earlier suggestion that GX\,349+2, Sco\,X-1 and 
GX\,17+2 are viewed at lower inclinations and Cyg\,X-2, GX\,5$-$1 and 
GX\,340+0 at higher inclinations (see also the discussion in Kuulkers et al.\ 
\cite{kko97}).

We note that broad power spectral peaks in the lower FB, near the NB/FB vertex, 
have also been reported for GX\,17+2 (FBO, Penninx et al.\ \cite{plt91}) and 
Cyg\,X-2 (Wijnands et al.\ \cite{wkk97}), with 
$\nu_{\rm peak}$$\sim$6--7\,Hz and FWHM$\sim$10\,Hz. However, these peaks
changed smoothly into typical NBO when the source moved from the lower
FB into the NB. Peaked broad-band noise components (LFN) in Z sources have been 
observed in Sco\,X-1 and GX\,17+2 (Dieters \&\ van der Klis 1997, Penninx et 
al.\ 1990, Kuulkers et al.\ 1997). However, these were observed {\it only} in 
the HB and upper NB, with peak frequencies of 2--3\,Hz, and were absent in the 
rest of the NB and in the FB. It is therefore unlikely the 
observed peaked noise in GX\,349+2 is a manifestation of LFN. 

The peaked noise in GX\,349+2 bears more resemblance in shape to the peaked 
noise components (referred to as high-frequency noise) seen in atoll sources 
in their so-called banana state (see Hasinger \&\ van der Klis 1989, van der 
Klis 1995b, and references therein). For example, the peaked noise in GX\,3+1 
found by Lewin et al.\ (\cite{lph87}, see also Hasinger \&\ van der Klis 1989), 
with centroid frequency, FWHM and 
fractional rms of $\sim$8\,Hz, $\sim$8\,Hz and $\sim$3.5\%, respectively, is 
very similar to that found by Ponman et al.\ (1988) and us. 
Moreover, the very-low-frequency noise (VLFN) in the bright atoll source 
GX\,13+1 was found to be anomalously strong (Hasinger \&\ van der Klis 1989), 
and similar to FB VLFN. This shows that some members of the atoll and Z sources
exhibit features of both classes and may suggest some overlap between the
classes. The recently discovered $\sim$60\,Hz QPO in GX\,13+1 (Homan et al.\
\cite{hxx97}) may be comparable to HB QPO and is even a stronger suggestion 
for the existence for such overlap.

So far, kHz QPO has been observed in the four Z sources Sco\,X-1
(van der Klis et al.\ 1996a, 1997b), GX\,5$-$1 (van der Klis et al.\ 1996b), 
GX\,17+2 (van der Klis et al.\ 1997a, Wijnands et al.\ 1997b) and Cyg\,X-2
(Wijnands et al.\ 1997c). We found only weak evidence of kHz QPO 
(at 2.6$\sigma$ confidence level), with a fractional rms, FWHM and centroid 
frequency of $\sim$1\%\ (2--30\,keV), $\sim$40\,Hz and $\sim$1020\,Hz,
respectively. We 
note that the expected frequency and fractional rms of the (higher frequency) 
kHz QPO in Sco\,X-1, GX\,17+2 and Cyg\,X-2 in the lower NB/FB are $>$1000\,Hz 
and $\sim$1\%, respectively. This is consistent with our observations.

\begin{acknowledgements}

This work was supported in part by the Netherlands Organization for
Scientific Research (NWO) and by the
Netherlands Foundation for Research in Astronomy (ASTRON)
under grants PGS 78-277 and 781-76-017, respectively.  
EK thanks the Astronomical Institute ``Anton Pannekoek'',
where part of the analysis was done, for its hospitality.

\end{acknowledgements}

\newpage

\section*{Figure captions}

{\bf Figure 1}: 
{\bf (a)} Light curve and {\bf (b)} hardness-intensity
diagram (HID) of GX\,349+2. The intensity is the source count rate in the 
2.0--19.7\,keV energy range. The hardness is defined as
the count rate ratio between 8.6--19.7\,keV and 6.0--8.6\,keV.
The start of the light curve is at 
UT~1997 May 2 18:44; data are at 16\,s time resolution. The data points in the 
HID are 64\,s averages. Typical error bars are shown.

{\bf Figure 2}: 
Leahy normalized power spectra (effectively 2.0--30\,keV), focussed on the 
low-frequency ($\la$100\,Hz) range, for the first
two continuous ({\bf a}, NB) and the last continuous ({\bf b}, FB) data 
segments. 

{\bf Figure 3}: 
Leahy normalized power spectrum (effectively 2.0--30\,keV) of the whole data 
set, focussed on the kHz QPO region. A possible QPO peak near 1020\,Hz is 
discernable at the 2.6$\sigma$ level.

\newpage

\begin{figure}
\psfig{file=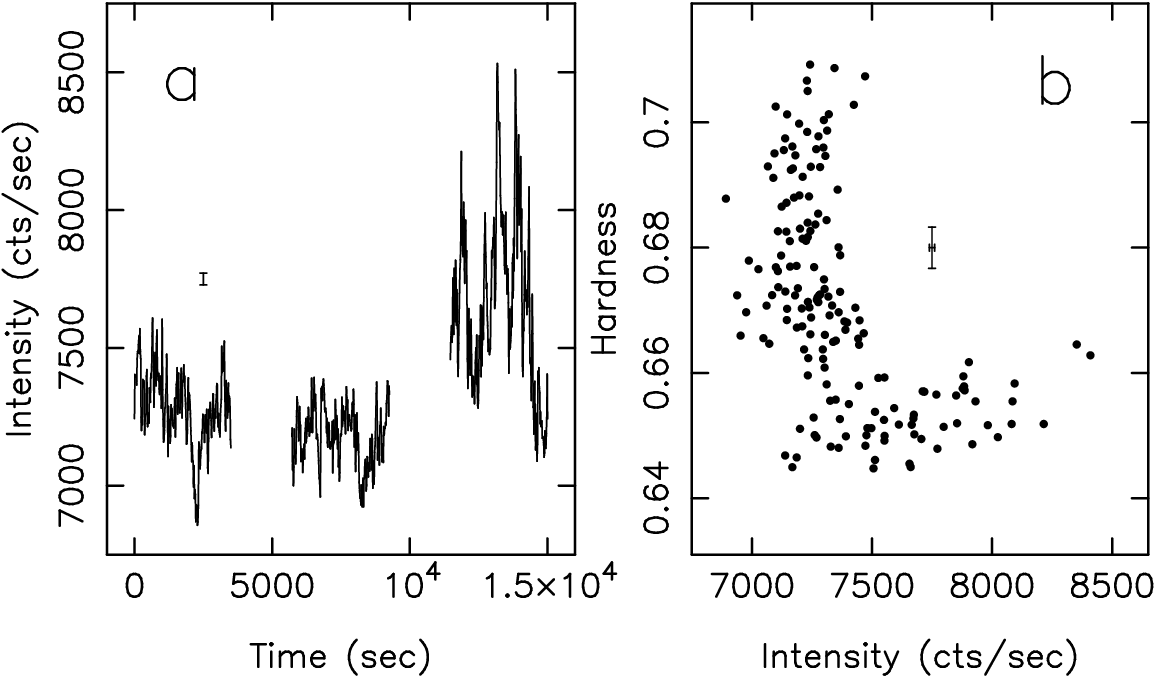,bbllx=23pt,bblly=368pt,bburx=591pt,bbury=707pt,height=10cm}
\caption{}
\end{figure}

\begin{figure}
\psfig{file=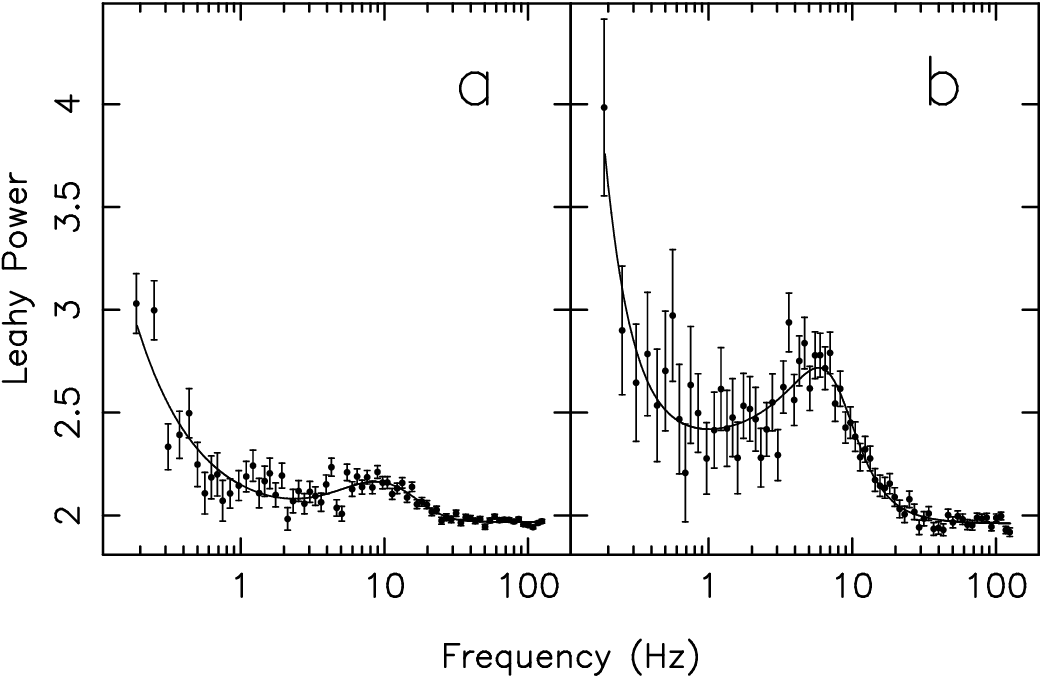,bbllx=53pt,bblly=368pt,bburx=567pt,bbury=707pt,height=10cm}
\caption{}
\end{figure}

\begin{figure}
\psfig{file=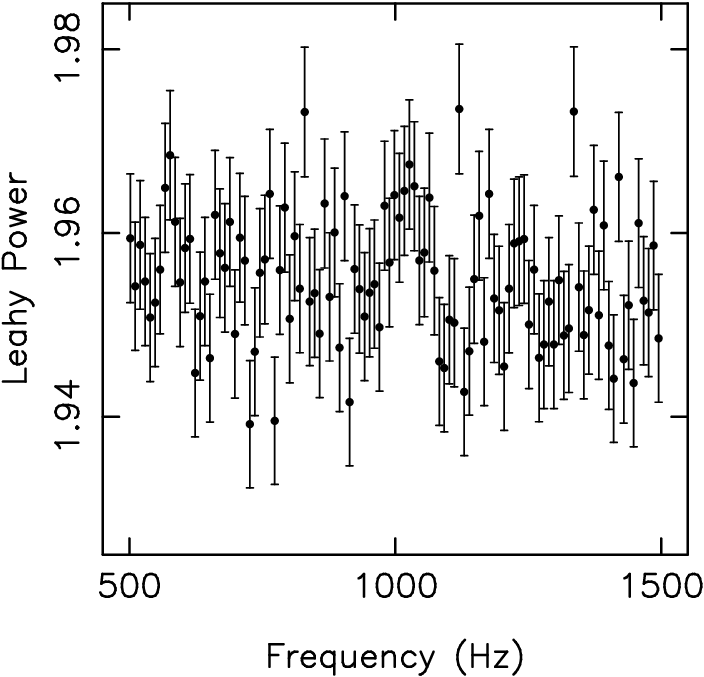,bbllx=25pt,bblly=368pt,bburx=370pt,bbury=707pt,height=10cm}
\caption{}
\end{figure}

\end{document}